\documentclass[a4paper]{scrartcl}

\pdfoutput=1
\usepackage[english]{babel}
\usepackage{authblk}

\usepackage{units}
\usepackage{caption}
\usepackage{xspace}
\usepackage[T1]{fontenc}
\usepackage{float}
\usepackage{latexsym}
\usepackage{url}
\usepackage{listings}
\usepackage{graphicx}
\usepackage{tikz}
\usepackage{tabularx}
\usepackage{array}
\usepackage{multirow}
\usepackage{fancyvrb}
\usetikzlibrary{petri}
\usetikzlibrary{positioning}
\usetikzlibrary{arrows}
\usepgflibrary{shapes}
\usetikzlibrary{decorations.pathreplacing}
\usetikzlibrary{calc}

\def\ltlexist\Diamond
\def\ltlall\oblong
\def\W{\mathrel{\mathbf{W}}}

\tikzstyle{transition} =[very thick, rectangle, draw, inner xsep=2mm, inner ysep=0.75mm]
\tikzstyle{vtransition}=[very thick, rectangle, draw, inner ysep=2mm, inner xsep=0.75mm]

\tikzstyle{place}=[circle, draw, minimum size=4ex]
\tikzstyle{every label}=[font=\sf\footnotesize]

\tikzstyle{pre}+=[>=stealth]
\tikzstyle{post}+=[>=stealth]
\tikzstyle{readarc}=[pre, >=*, shorten <=0pt]
\tikzstyle{prio}=[draw, ->, orange, >=stealth, shorten >=1.25pt, shorten <=1.25pt, densely dashed]
\tikzstyle{poids}=[font=\scriptsize\sf]

\tikzstyle{action}=[rectangle, draw, color=black!80, thin, dotted, inner xsep=0.4ex, inner ysep=0.1ex]

\newcommand{\tin}[1]{\textsf{\small #1}}
\newcommand{\code}[1]{\texttt{\small #1}}

\definecolor{mygreen}{rgb}{0,0.6,0}
\begin{document}

\title{Latency Analysis of an Aerial Video\\ Tracking System Using
  Fiacre and Tina}

\author[1,2]{Silvano Dal Zilio}
\author[1,2]{Bernard Berthomieu}
\author[1,3]{Didier Le Botlan}
\affil[1]{CNRS, LAAS, F-31400 Toulouse,  France}
\affil[2]{Univ de Toulouse, LAAS, F-31400 Toulouse, France}
\affil[2]{Univ de Toulouse, INSA, LAAS, F-31400 Toulouse, France}
\date{}
\maketitle
\begin{abstract}
  We describe our experience with modeling a video tracking system
  used to detect and follow moving targets from an airplane. We
  provide a formal model that takes into account the real-time
  properties of the system and use it to compute the worst and
  best-case end to end latency. We also compute a lower bound on the
  delay between the loss of two frames.

  Our approach is based on the model-checking tool Tina, that provides
  state-space generation and model-checking algorithms for an
  extension of Time Petri Nets with data and priorities. We propose
  several models divided in two main categories: first Time Petri Net
  models, which are used to study the behavior of the system in the
  most basic way; then models based on the Fiacre specification
  language, where we take benefit of richer data structures to
  directly model the buffering of video information and the use of an
  unbounded number of frame identifiers.
\end{abstract}


\section{Introduction}
\label{sec:introduction}

We describe our experience with modeling a video tracking system used
on board an aircraft to detect and follow moving targets on the
ground. This industrial case study has been submitted as a
verification challenge during the 6th International Workshop on
Analysis Tools and Methodologies for Embedded and Real-time
Systems. We propose an answer to the first challenge of this use
case~\cite{waters}. Our solution is based on the use of a
model-checking tool for an extension of Time Petri Nets with data and
priorities. The models used in this study are available online at
\url{http://www.laas.fr/fiacre/examples/videotracking.html}.


The purpose of the video tracking system is to detect objects of
interest inside a stream of images coming from a video camera (such as
a moving vehicle for example); to estimate their future position; and
to control the camera in order to track these objects over long
periods of time. This system includes two high-level tasks: (1) a
\emph{video processing system} that process camera frames in order to
embed tracking information and, ultimately, to display them on a
monitor ; and (2) a \emph{tracking and control system} that performs
the motion prediction tasks and control the orientation of the camera
based on the aircraft sensors data (position, direction and speed,
etc..)

We focus on the real-time properties of the video frame processing
system (the first challenge) and do not consider the scheduling issues
raised by the interaction between the two high-level tasks. The {video
  frame processing} system is essentially a graphics pipeline, with
four processing stages working concurrently but connected in
series. The first stage of this pipeline is to fetch a new image frame
from the camera; while the final stage consists in sending the
processed frames to the display.


\begin{figure}[!t]
\centering
\includegraphics[width=0.8\linewidth]{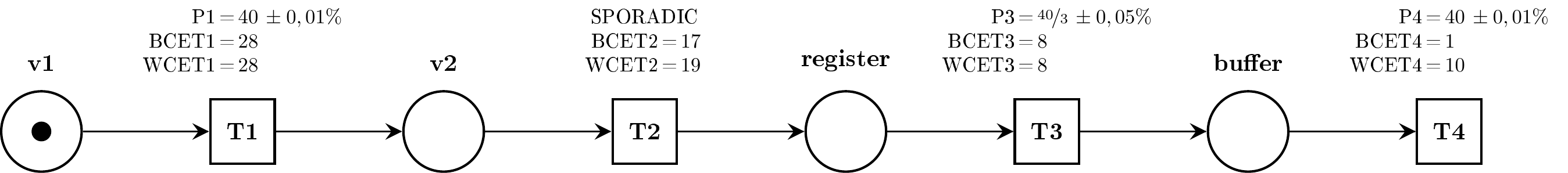}
\vspace{1em}
\hrule
\vspace{1em}
\includegraphics[width=\linewidth]{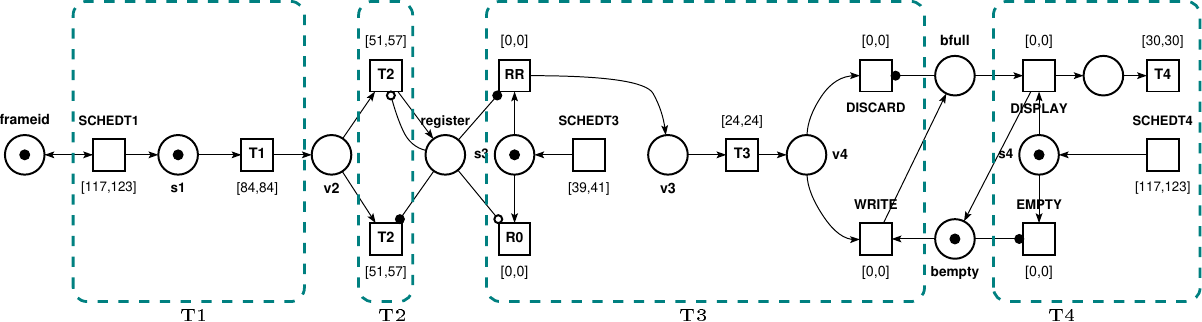}
\caption{The Video Frame Processing Pipeline: schematic view (above)
  and initial TPN model (below)}
\label{fig::system}
\end{figure}

The main goal of the challenge is to compute the \emph{latency} of the
system, that is the best and worst-case delay from start (reception of
a new frame) to finish (output to the display). Several factors
complicate the problem. First, each step of this pipeline have
different---and sometimes varying---computing times and are scheduled
with different periods. Therefore some amount of buffer storage is
needed and the occurrence time of the frames arrival at each stage of
the pipeline cannot be defined with an analytical formula. Next, there
is a small amount of uncertainty on the periods of the tasks. Whereas
the stages in the pipeline are in sync when the periods are exactly
those given in the specification, even a small divergence (accumulated
over several periods) can lead to an arbitrary offset between
tasks. We show that, in this case, video frames can be lost. More
precisely, some frames will never reach the last stage of the pipeline
because they will never be written in the display buffer; frames are
discarded when the buffer is full. Finally, even though there is only
a finite number of different frames in the processing pipeline at any
given moment, the system deals with a potentially unbounded number of
\emph{frame identifiers}. In our approach, we use different modeling
techniques to abstract away this unbounded number of identifiers. As a
consequence, we can solve our problems using model-checking techniques
developed for finite-state systems.

Our approach is based on the model-checking tool Tina~\cite{tina8},
that provides state-space generation and model-checking algorithms for
an extension of Time Petri Nets with data and priorities. We propose
several models following two levels of accuracy. First, we use Time
Petri Net (TPN) in order to study the behavior of the system in the
most basic way. In this case, we restrict ourselves to the case where
the buffer is of size one. Second, we use the Fiacre specification
language~\cite{Berthomieu-ERTS08}, where we take benefit of richer
datatypes to model different sizes of buffers and an unbounded number
of frame identifiers more simply. Both the extended TPN and the Fiacre
models can be used for analysis by Tina.


\section{Description of the Frame Processing System}

We start by describing the architecture of the Frame Processing System
pipeline (FPS) using a simple diagram, see Fig.~\ref{fig::system}
(above). The diagram outlines the intermediate data values (the
``channel'' v2, a register and a buffer) that are used as storage
between each stage. The system is composed of four different functions
that are each mapped to a unique task, T1 to T4, and executed on
separate hardware:

\noindent\textbf{Task T1} is in charge of pre-processing the video frames
coming from the camera. Each frame is assigned a unique identifier,
that we call an \emph{id} in the rest of the paper. New frames arrive
strictly periodically with a period P1 that is a (constant) value in
the range $40 \pm 4.10^{-3}$. Once a new id arrives, T1 outputs a
preprocessed frame, with the same identifier, after \unit[28]{ms}.  To
compute the latency, we can abstract away the role of the camera and
simply consider that T1 is a periodic task of period P1. Also, we do
not need to model the actual image payload; we can simply consider
that ids are the only data exchanged between tasks.

\noindent\textbf{T2} is a  sporadic task, triggered by the arrival of a frame
at its input (v2 in our diagram). The execution time of T2 may vary
but is always in the interval $[17,19]$. The task overwrites its
output register content when it is finished. The specification does
not fix the initial values of the intermediate data values. We
consider that v2, the register and the buffer are all initially
empty. To this end, we will sometimes use a ``dummy'' id (denoted
NIL) to distinguish the initial value.

\noindent\textbf{T3} is a periodic task, with a period P3 in the range
$\displaystyle\nicefrac{40}{3} \pm \nicefrac{2}{3}.10^{-2}$. When
scheduled, T3 fetches an id in the register and outputs its result in
a buffer after \unit[8]{ms}; the result is not written---it is simply
discarded---when the buffer is full. (We assume that T3 does not
execute when the register is empty.) Since P1 is roughly three times
bigger than P3, task T3 will often process the same register value
more than once.

\noindent\textbf{Task T4} is periodic, with a period P4 that is in the
same range than P1. When T4 is scheduled, and if the buffer is not
empty, the last id is dequeued. In this case, the task takes
\unit[10]{ms} to compute its output for the display. Due to
uncertainties on the hardware clocks, the periods of T1 and T4 may be
slightly different but cannot deviate by more than $0,02\%$. This very
small difference between the timing constraints of the system is
typically a source of combinatorial explosion when
model-checking. Indeed, this means that we need roughly $10\,000$
periods before coming back to a ``previously explored'' configuration
of the system. For this reason, we will use the amount of errors on
the periods as a parameter of the system and study its influence on
the complexity of our approach.

Since our state-space generation algorithms are based on dense-time
techniques, our approach is insensitive to the choice of a scaling
factor. Therefore, multiplying the timing constants by a fixed amount
does not change the results or the performance of our tools. In all
our experiments, we will mainly use three sets of timing constraints
summarized in the following table. The last row gives the scaling
factor (we choose the same values for T1 and T4): SPEC are the values
obtained from the specification; WIDE are values obtained with an
error in the order of 1\%;
EXACT are values without errors.\\

\begin{center}
  {\noindent\begin{tabular}[h]{|l|p{7.6em}|p{6.6em}|p{8em}|} \hline
              & T1 = T4 & T3 & scale / error\\
              \hline
              SPEC  & [119\,988, 120\,012] & [39\,980, 40\,020] & 3\,000 / 0.01\%\\
              \hline
              WIDE  & [117, 123] & [39, 41] & 3 / 2--3\%\\
              \hline
              EXACT  & [120, 120] & [40, 40] & 3 / 0\%\\
              \hline
            \end{tabular}}
\end{center}

\section{A Simple Time Petri Net  Interpretation}
\label{sec:first-interpr-using}

We provide a first set of solutions to the challenge based on Time
Petri Nets (TPN). For reasons of brevity, we assume a basic knowledge
of the semantics of Petri Nets and we refer the reader to
e.g.~\cite{BV07} for an introduction to semantics of TPN.

Time Petri Nets share the same graphical representation than standard
Petri Net. In addition, a transition $t$ can be decorated with a time
interval, $I(t)$. Informally: a transition $t$ is enabled if there is
enough tokens in the places connected to it; time can elapse, at the
same rate, on all the transitions that are continuously enabled;
finally, $t$ can fire if it has been enabled for a time $\theta$ such
that $\theta \in I(t)$. In particular, transitions associated to the
interval $[0,0]$ should be fired immediately.
\subsection{Description of the TPN model}

We give a very simple interpretation of the frame processing system in
Fig.~\ref{fig::system}. This net uses the WIDE set of timing
constraints, therefore all timing constraints are multiplied by a
factor of $3$. The model was generated using the tool \code{nd}, which
is an editor and simulator for TPN that is part of Tina (see the file
\code{initial-wide.ndr} in our source files). For documentation, we
have superimposed dashed boxes to the model in Fig.~\ref{fig::system}
in order to stress which parts of the TPN is related to which task.

Our modeling choices are quite simple. Actually, our main goal is to
provide an interpretation that is as close as possible to the
specification; we avoid possible optimizations in order improve
readability of the model.  We have two transition labels,
\tin{SCHEDT}$i$ and \tin{T}$i$, for every periodic task T$i$ (with $i$
in $\{1,3,4\}$). Transition \tin{SCHEDT}$i$ is used to start
(schedule) the task periodically. When fired, it places a token in the
place \tin{s}$i$ that will stay for the duration of the task
execution. The role of \tin{T}$i$ is to model the end of the execution
and the associated side-effect. For instance, transition \tin{T3}
models the insertion in the buffer and \tin{T4} models sending the
final frame to the display.

The one-place buffer between T3 and T4 is modeled using two places,
where a token indicates the current state; empty or full. When
\tin{T3} fires, and if the buffer is empty, then it becomes full after
a zero delay (transition \tin{WRITE}); otherwise the ``write'' is
discarded. Task T3, the most complex component of our model, has two
additional transitions, \tin{R0} and \tin{RR}, that model reading from
the register when it is either empty or full. In this case, we use
inhibitor arcs (displayed with a $\mbox{---}\circ$ arc) that block a
transition when a token is present. We also use read arcs
($\mbox{---}\bullet$) that can test the presence of a token without
disturbing the enabled transitions.  


The initial marking of the net states that every periodic task are
scheduled at time $0$ and that the buffer and register are empty.

\subsection{Behavioral Verification with Tina}
\label{sec:behav-verif-with}

Tina~\cite{tina8}, the TIme Petri Net Analyzer, provides a software
environment to edit and analyze Time Petri Nets and their
extensions. The core of the Tina toolset is an exploration engine,
called \code{tina}, used to generate state space abstractions that can
be later exploited by dedicated model checking and transition system
analyzer tools. (Most of the components in Tina are command line tools
and are available on the most common operating systems.)  In our
experiments, we use the tool \code{sift} instead of
\code{tina}. \code{Sift} is a specialized version of \code{tina} that
supports on the fly verification of reachability properties. It offers
less options than \code{tina} but is typically faster and requires
considerably less space when dealing with large models.

State space abstractions are vital when dealing with timed systems,
such as TPN, that have in general infinite state spaces (because we
work with a dense time model).  Tina offers several abstract state
space constructions that preserve specific classes of properties like
absence of deadlocks, reachability of markings, linear time temporal
properties, or bisimilarity.  In the case of the FPS, most of
the requirements can be reduced to safety properties, that is,
checking that some ``bad state'' cannot occur. In this case, we do not
need to generate the whole state class graph of the system and we can
use ``more aggressive'' abstractions. 

Tina implements two main state-space abstraction methods, a default
method that preserves the set of states and traces of the system, and
a method that preserves the states but over-approximate the set of
traces. This second method can be used in Tina with the command line
options \code{-M}; it is often much more efficient than the default
exploration mode

To give a rough idea of the complexity of our initial model, it is
possible to generate its exact state class graph and, for example, to
check that the net is 1-safe (that the number of tokens in each place
is always at most $1$). This is done using the flag \code{-b} in Tina.
The property is true and can be checked in less than \unit[10]{s} on a
modern laptop with 4GB of RAM. The same property can be checked almost
instantaneously ($\leq$ 0.01s) with the option \code{-M}. With the
SPEC timing constraints, the same computation with option \code{-M}
takes \unit[28.8]{s} and use a peak memory of \unit[110.7]{MB}.

We can already make a few ``sanity checks'' using our model. For
instance, since the model is 1-safe, we prove that a task cannot
receive a new id while it is still executing; otherwise one of the
place s$i$ would hold more than one token.  We can also use this model
to check the impact of the initial conditions on the behavior of the
system. For instance, if the pipeline starts with a frame in the
register (there is a token in place \tin{register} initially) then we
can prove that the buffer cannot be empty when T4 is scheduled; that
is the transition \tin{EMPTY} is never enabled. As a consequence,
except for a first few periods at the start of the system, we know
that the execution of T4 will always takes \unit[10]{ms}.

\begin{figure}[htp]
\centering
\includegraphics[width=\linewidth]{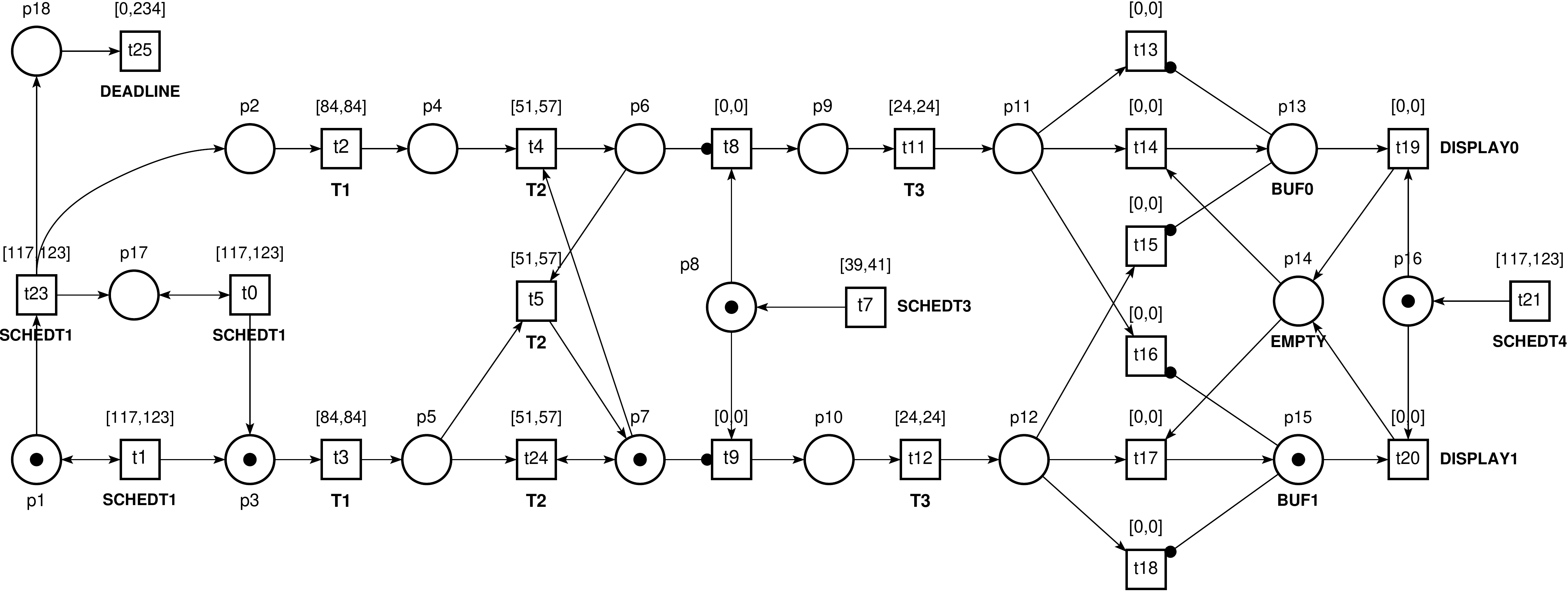}
\caption{A TPN model for computing the best-case latency (using two
  frame identifiers)}
\label{fig::system2}
\end{figure}

On the other hand, we cannot use the model in Fig.~\ref{fig::system}
to compute the latency or to check that frames can be lost. The main
reason is that we cannot distinguish two different id by looking at
the marking (every id is represented as a token). We can easily change
the model to consider a finite number of id by creating copies of each
place (one copy for every possible value of id). We give an example
of the TPN with two different id in Fig.~\ref{fig::system2}: a token
in place \code{p3} (resp.  \code{p2}) means that we generate a frame
with id $1$ (resp. $0$). The net has an extra transition, \code{t25},
that can fire at most $\Delta = 234$ after a frame with id $0$ is
generated. At the other end of the pipeline, this frame is send to the
display if transition \code{t19} fires. Therefore, to test if the
best-case latency is less than $\Delta$, it is enough to prove that we
cannot fire \code{t19} while there is still a token in place
\code{p18}. This is a reachability property that can be checked in
Tina with the option \code{-f}. The following command executes in
\unit[0.4]{s} on the model with the WIDE timing constraints (and in
\unit[101.6]{s} without option \code{-M}):\par {\code{sift -M -f "t19
    => -p18" initial-wide.ndr}}\par

Another problem is to model a buffer with $n$ places. It is possible
to encode this data structure using a TPN but it can quickly become
awkward when the value of $n$ grows big. Also, we would like to
compute the minimal number of different id necessary to check our
properties. For this reason we propose a high-level model of the Frame
Processing System using the Fiacre specification language.

\section{An Interpretation Using Fiacre}

Fiacre is a specification language designed to represent both the
behavioral and real-time aspects of systems
({\url{http://www.laas.fr/fiacre/}}). The language comes equipped with
a set of dedicated tools, such as \code{frac}, a compiler from Fiacre
to a format handled by Tina. In a nutshell, a Fiacre specification is
compiled into an extension of TPN where transitions can test and
modify a set of data variables. Therefore Fiacre provides a
high-level, compositional syntax for defining TPN. A formal definition
of the language is given in~\cite{BERTHOMIEU-FIACRE07}.

Fiacre programs are stratified in two main notions: \emph{processes},
which is basically a notation for states machines, and
\emph{components}, which describe a system as a composition of process
instances. Components can also be nested.

The language supports two of the most common coordination
paradigms---communication through shared variable (shared-memory) and
synchronization through synchronous communication ports
(message-passing)---and allows to define timing constraints on the
transition of a system. We give an example of process declarations in
Fig.~\ref{fig::process}. This code corresponds to the tasks T3 and T4
of the FPS (we still use the WIDE timing constraints). We follow the
same philosophy in our Fiacre specification than with our TPN
solution; we provide a model that is as close as possible to the
specification and we favor readability over optimization of the model.

{\lstset{basicstyle=\scriptsize\ttfamily,
  identifierstyle=\ttfamily, 
  keywordstyle=\ttfamily\bfseries\color{blue},
  captionpos=b,  
  commentstyle={\color{mygreen}\itshape},
  frame=none,
  numbers=none
}
\begin{figure}[htbp]
  \centering
  {
\begin{lstlisting}[language=FIACRE]
process T3 (&register: id, &buffer : mbuff) is
   states set, get

   var v3 : id := NIL
   
   from set
      wait [24,24];
      buffer := insertbuff (buffer, v3)
      to get

   from get
      wait [15,17];
      v3 := register;
      to set

//----------------------------------------------------  

process T4 (&buffer : mbuff) is
   states set, get, displayed 

   from set
      wait [0, 0];
      if (empty buffer) then
	 to get
      else
	 case first(buffer) of
	    MAGIC  -> to displayed
	 |  any    -> buffer := dequeue buffer; to get
	 end
      end
      
   from get
      wait [117, 123];
      to set

//----------------------------------------------------

component C is
   var v1 : id := NIL, v2 : id := NIL,
      register : id := NIL,
      buffer : mbuff := {||}

      par
	 camera (&v1)
      || T1 (&v1, &v2)
      || T2 (&v2, &register)
      || T3 (&register, &buffer)
      || T4 (&buffer)
      end
      
\end{lstlisting}}
  \caption{Excerpt from our Fiacre solution}
  \label{fig::process}
\end{figure}
\begin{figure}[htbp]
  \centering
  {
\begin{lstlisting}[language=FIACRE]
const CAPACITY : nat is 3

type id is union
   NIL
|  FRAME of nat
|  MAGIC
end
   
type mbuff is queue CAPACITY of id

function insertbuff (q: mbuff, f: id) : mbuff is
   var tmp: mbuff := q
   begin
      if (full q) then
	 return q
      end;
      while not (empty tmp) do
	 if (f = first tmp) then
	    return q
	 end;
	 tmp := dequeue tmp
      end;
      return enqueue(q, f)
   end
\end{lstlisting}}
  \caption{Types, constants and functions}
  \label{fig::types}
\end{figure}}

\subsection{Description of the Fiacre model}

The whole system is defined in the component \code{C} (see
Fig.~\ref{fig::process}), where we state that the system is the
parallel composition of five different process instances,
\code{camera} to \code{T4}. Components are the unit for process
instantiation and for declaring ports and shared variables. The
declaration of \code{C} states that the buffer is initially empty and
that all the other variables are set to the id \code{NIL}.

Fiacre is strongly typed, meaning that type annotations are exploited
in order to guarantee the absence of unchecked run-time errors. The
language offers a large choice of data structures, such as natural
numbers, arrays, records, \dots Type declarations for our model are
listed in Fig.~\ref{fig::types}. The register is a variable holding
values of type \code{id}.  A value of type \code{id} can be of three
kind: the constant \code{NIL}; a ``common'' frame with index $i$,
denoted \code{FRAME}($i$); a special constant, \code{MAGIC}, that is
the frame id that we monitor when computing the latency. The buffer,
of type \code{mbuff}, is a fifo queue of length \code{CAPACITY} that
can hold values of type \code{id}. The fifo queue is a primitive type
in Fiacre, but the insertion policy of the buffer in the FPS is not
standard: the id is discarded when the buffer is full; otherwise we
add a frame id only if it is not already present in the queue. To
model this behavior, we define a dedicated insertion function in
Fiacre (see the declaration of \code{insertbuff} in
Fig.~\ref{fig::types}).

A Fiacre process is defined by a set of parameters and {control
  states}, each associated with a set of \emph{complex transitions}
(introduced by the keyword \code{from}). Complex transitions are
expressions that declare which transitions may fire and how variables
are updated. Expressions are built from classical programming
languages constructs (assignments, conditionals, pattern-matching,
\dots); non-deterministic constructs (such as external choice); jump
to a state (\code{to}); etc. As an example, we describe the behavior
of process \code{T3}. (The other processes are roughly similar.)
Process \code{T3} can access two shared variables, \code{register} and
\code{buffer}. One can remark that the parameters of each process are
exactly those defined in our schematic view of
Fig.~\ref{fig::system}. When \code{T3} has been in the state
\code{get} for a duration in $[15,17]$, it copies the value of the
register in a local variable, \code{v3}, and move to the state
\code{set}. After waiting another $24$, it then tries to insert the
value of \code{v3} in the buffer and move back to \code{get}. This
models the behavior of a periodical task, with period $40$ and jitter
$1$, that writes its into the buffer with a delay of $24$.

\subsection{Latency Analysis of the Fiacre Model}

Figure~\ref{fig::camera} gives the code for process \code{camera},
that is in charge of generating new frames. This process writes
periodically a new id in the variable \code{v1}; it will write only
once, non-deterministically, the special value \code{MAGIC} then go to
state \code{stop}. In \code{camera}, we use the user-defined function
\code{nextid} to compute, and reuse, the first frame identifier that
does not occur in the system. In this way, we can prove that we need
only a finite number of values of the form \code{FRAME}($i$) without
bounding a priori the value of $i$.

On the other end of the pipeline, values of type \code{id} are
consumed by \code{T4} (see Fig.~\ref{fig::process}). Process \code{T4}
will move to the state \code{displayed} as soon as it spots the
special id \code{MAGIC}. We can use this state to compute the
end-to-end latency of a frame; we only need to check the time between
the creation of our special frame (when \code{v1} equals \code{MAGIC})
to its eventual display (when \code{T4} enters state
\code{displayed}). In practice, this can be done by adding an
``observer'' process to the component \code{C} that monitors the time
between these two events.
{\lstset{basicstyle=\scriptsize\ttfamily,
  identifierstyle=\ttfamily, 
  keywordstyle=\ttfamily\bfseries\color{blue},
  captionpos=b,  
  commentstyle={\color{mygreen}\itshape},
  frame=none,
  numbers=none
}
\begin{figure}[htbp]
  \centering
  {
\begin{lstlisting}[language=FIACRE]
process camera (&v1: id, ...) is
   states send, stop

   from send
      wait [117,123];
      select
	 v1 := FRAME(nextid(v1, ...));
	 to send
      [] v1 := MAGIC;
	 to stop
      end

   from stop
      wait [117,123];
      v1 := FRAME(nextid(v1, ...));
      to stop

process observer(&v1 : frameid) is
   states start, deadline, stop

   from start
      wait [0,0];
      on (v1 = MAGIC);
      to deadline

   from deadline
      wait [0,234]; // Delta = 234
      to stop
      
//--------------------------------------------------

function isFrame (u : id, F : nat) : bool is
   begin
      case u of
	FRAME(F) -> return true
      | any -> return false
      end
   end
   
//--------------------------------------------------

property minlatency is ltl [] ((C/5/state displayed)
                                 => (C/6/state stop))

property max is ltl [] not (C/1/value isFrame(v1,5))
      
\end{lstlisting}}
  \caption{The \code{camera} process with Fiacre properties}
  \label{fig::camera}
\end{figure}}

We give an example of observer in Fig.~\ref{fig::camera}. Process
\code{observer} enters the state \code{deadline} as soon as the guard
\code{(v1 = MAGIC)} is true. The process cannot stay in this state for
more than $\Delta = 234$. As a consequence, the ``latency'' of the
frame \code{MAGIC} is bigger than $\Delta$ if and only if we can reach
a state such that \code{T1} is in the state \code{displayed} while
\code{observer} is not yet in state \code{stop}. We can express this
property directly in the Fiacre code using a \code{property}
declaration; we give two examples of property in
Fig.~\ref{fig::camera}. Therefore, to compute the best-case latency,
we need to check the property \code{minlatency} for several values of
$\Delta$ and select the ``first value'' such that the property is
false. A similar approach can be used to compute the worst-case
latency and the minimal time between the generation of two lost
frames. We give the results of our analysis in the next section.

The second property in Fig.~\ref{fig::camera}, \code{max}, can be used
to check that it is not necessary to consider more than $5$ different
id when the buffer is of size $3$; i.e. the variable \code{v1} is
always different from \code{FRAME(5)}. 



\section{Experimental Results}
\label{sec:experimental-results}

The models used in our experiments are available online at:
\url{http://www.laas.fr/fiacre/examples/videotracking.html}.  Most of
the models were developed in the course of one week, but we were able
to have our first TPN model in about one hour. The TPN and Fiacre
models where done concurrently by two different people (we obtain the
same values with the two approaches). This development speed can be
explained by the fact that we can easily model-check our examples; a
modification to the model can be tested in a few seconds. Also, the
size of the models are quite reasonable and therefore it was possible
to use our simulators to understand the behavior of the system and to
analyze the counter-examples returned by the model-checker.

Most of the time was spent refactoring our models in order to simplify
their presentation for this paper. We also spent some time analyzing
examples of ``execution traces'' to understand which scenarios led to
the loss of a frame or to the best-case latency. For instance, we
found that the scenarios for the worst-case latency is almost the same
than for a frame loss, which is not surprising as an afterthought; the
``slowest'' frame is the one that was almost lost. We give the values
computed in our experiments using the different timing constraints
(the result are converted into \unit{ms}) for two possible sizes of
the buffer, $n = 1$ and $n = 3$. Column BTW gives the minimal time
between two lost frames; it is easy to understand that BTW is
necessarily a multiple of the period P1.

Our models are not totally faithful to the specification. In
particular, we use an interval to model the uncertainty on the
period. This means that, inside the same execution trace, the periods
of a task may vary. (It would be possible to have ``an exact'' model
using an extension of TPN with stopwatches, but this is much more
costly in terms of performances.) As a result, we obtain an
over-approximation of the possible behaviors: this gives a lower-bound
when computing the minimal latency and an upper-bound for the maximal
latency. The same observation is true when we over-approximate the
possible error on the periods; results obtained in the case WIDE will
give lower and upper-bound for the case SPEC (which bounds the optimal
values).

Another possibility is to compute the latencies by iterating over a
sample of values for P1, P3 and P4. If we denote by
$\code{MIN}_{\code{FIX}}$ and $\code{MAX}_{\code{FIX}}$ the best
latencies value obtained on a sample of periods values and by
\code{MIN} and \code{MAX} the exact best and worst-cases we have the
following relations. (The relations for \code{BTW} are the same than
for \code{MIN}.)
\[
\begin{array}{c@{\,\le\,}c@{\,\le\quad}c@{\quad\le\,}c@{\,\le\,}c}
  \code{MIN}_{\code{WIDE}} & \code{MIN}_{\code{SPEC}} & {\code{MIN}}
  & \code{MIN}_{\code{FIX}} & \code{MIN}_{\code{EXACT}}\\[0.5em]

  \code{MAX}_{\code{EXACT}} &  \code{MAX}_{\code{FIX}} & \code{MAX} & \code{MAX}_{\code{SPEC}}  & \code{MAX}_{\code{WIDE}}
\end{array}
\]


\begin{figure} \centering
  {
\begin{tabular}[h]{|l||r|r|l|}
   \hline
   $n = 1$ & MIN & MAX & BTW\\
   \hline
   \hline
   EXACT  & \unit[130]{ms} & \unit[130]{ms} & n.a. \hfill (no loss)\\
   \hline
   WIDE  & \unit[88]{ms} & \unit[147]{ms} & \unit[78]{ms} \hfill ($2$
   periods)\\
   \hline
   SPEC  & \unit[89.65]{ms} & \unit[145]{ms} & \unit[79.99]{ms}
   \quad($2$ periods)\\
   \hline   
   FIX  & \unit[90]{ms} & \unit[144]{ms} &  \unit[79.99]{ms}
   \quad($2$ periods)\\
   \hline
 \end{tabular}}
\end{figure}

\begin{figure} \centering
  {
\begin{tabular}[h]{|l||r|r|l|}
   \hline
   $n = 3$ & MIN & MAX & BTW\\
   \hline
   \hline
   EXACT  & \unit[130]{ms} & \unit[130]{ms} & n.a. \hfill (no loss)\\
   \hline
   WIDE  & \unit[88]{ms} & \unit[229]{ms} & \unit[429]{ms} \hfill ($11$ periods)\\
   \hline
   SPEC  & \unit[89.65]{ms} & \unit[225]{ms} & > \unit[1.4]{s} \hfill($35$
   periods)\\
   \hline
   FIX  & \unit[90]{ms} & \unit[223.33]{ms} &  < \unit[163]{s} \quad($4085$
   periods)\\
   \hline
 \end{tabular}}
\end{figure}

Our experiments show that there is no loss of frames when the periods
P1 and P4 are equal (even if P3 is not exact). We also observe that
increasing the size of the buffer does not reduce the best-case
latency but can aggravate the worst-case.

We have not been able to compute the value of BTW for the 3-place
buffer in the case SPEC (we stop computations that takes more than an
hour). We can provide a lower-bound for $\code{BTW}$, obtained using
finer intervals than with \code{WIDE}. Based on the observations made
during the computation for the \code{FIX} case, we believe that the
``exact value'' of $\code{BTW}$ could be much bigger than this bound
(it appears that $\code{BTW}$ is inversely proportional to the
difference between periods, $|\mathrm{P1} - \mathrm{P4}|$). Our best
upper-bound for \code{BTW} was obtained using an optimized model; we
find a bound of approx. $4\,000$ periods for \code{FIX} using the
values P1 = \unit[39.996]{ms}, P3 = \unit[13.333]{ms} and P4 =
\unit[40.004]{ms}. This is the only question in the challenge for
which we need to use a ``more clever approach'' than the simple,
straight solution that was enough up to now. Nonetheless, more work is
still needed in order to obtain a satisfactory value for \code{BTW}.


\section{Conclusion}


We have used a real-time model-checker to compute the maximal
(worst-case) and minimal (best-case) end to end latency of the Frame
Processing System, as well as to find a lower bound on the delay
between the loss of two frames. Therefore, instead of using
model-checking for validation, as is usual, we use it as a tool for
architecture exploration. 

This case study is interesting for several reasons. First, it is
well-suited for component-based modeling languages (since the
description is highly modular) and it is a good example for real-time
verification methods (since the specification has plenty of timing
constraints). Also, this case study provides a good motivation for the
use of high-level data structures in a specification language. In our
Fiacre models, for instance, we use a queue of identifiers with a
dedicated insertion function to elegantly model the buffer's behavior.

Finally, our experiments show that the very low level of inaccuracy on
the periods---typically a hundredth of a percent in the
specification---is a major source of combinatorial explosion. (We are
in a case where bigger errors leads to better performances). In this
context, we show that we can use dedicated state space abstraction
techniques (in this case the option \code{-M} of Tina) in order to
solve problems where the most general approach fail to scale up. This
stress the importance to provide a full verification toolbox that
gives access to a range of optimizations and modeling help.

\newpage


\bibliographystyle{plain}
\bibliography{waters}

\end{document}